\documentclass{elsart}
\usepackage{amssymb}
\usepackage{amsmath}
\usepackage[dvips]{graphicx}

\journal{:\bf \quad  C. R. Acad. Sci. Paris, t. 328,   II b, p.
151-157,
2000.\qquad\qquad\qquad\qquad\qquad\qquad\qquad\qquad\qquad\qquad\qquad
}
\input{tcilatex}
\begin{document}

\begin{frontmatter}

\title{Bubble Number in a Caviting Flow}
\author{Henri Gouin}
\ead{henri.gouin@univ-cezanne.fr}

\address {
C.N.R.S.  U.M.R. 6181 \&  Universit\'e d'Aix-Marseille \\ Case 322,
Av. Escadrille
 Normandie-Niemen, 13397 Marseille Cedex 20 France}
\author{Laurent Espanet}
\address {
CEA, Centre de Cadarache DRN/DEC/SH2C/LHC, \\ 13108 Saint Paul Lez
Durance, France}

\begin{abstract}
Cavitation is a general phenomenon of the fluid flows with
obstacles. It appears in the cooling conduits   of the fast nuclear
engines. A model of this phenomenon using the theory of Laplace and
a common
 non-convex energy  for  the  liquid and vapour bulks is proposed. This model makes it possible
 to determine a higher limit of the density of bubbles (a number of bubbles per unit of volume in
 the flow). The maximum intensity of cavitation is associated with the mechanical and thermal
characteristics of the fluid flow.
\end{abstract}

\begin{keyword} Gas/liquid flows;  Bubble formation; Bubble
dynamics; Cavitation.
  \PACS 47.55.Ca; 47.55.db; 47.55.dd; 47.55.dp
\end{keyword}

\end{frontmatter}

\section{Introduction}

The Laplace theory applied to a closed tank allows us to obtain an
equilibrium equation of a liquid in presence of vapour bubbles. We
analytically prove that the total number of bubbles cannot exceed a maximum
value. We extend the model to the mixture of fluid and gas in the case of
steady flows without viscosity: the number of bubbles transported by the
current admits an upper limit. \newline
It was observed in experiments that the number of bubbles created in a
cavitating flow depends on the quality of the carrying fluid. It is well
known that an injection of gas bubbles does not affect this upper limit. The
number of particles and microscopic gas bubbles is a significant cause of
the intensity of cavitation [1-3]. Unfortunately, this parameter is
generally inaccessible to measurement. \newline
The practical case is associated with the flows existing in coolant circuits
of fast nuclear engines; the used cooling agent consists of a mixture of
sodium and argon. Sodium in liquid form or vapour is called "fluid". Argon
in form of gas, possibly dissolved, is a neutral constituent at a
temperature much higher than its critical value. It is called "gas". It is
not possible to analyze the quality of the mixture. To estimate a maximum
number of bubbles produced by a possible phenomenon of cavitation is very
significant. It is an essential data of the nuclear security.

\section{Simplified model of cavitating flow}

We consider an isothermal permanent two-phase flow made up of fluid (for
example sodium at the temperature of $400 {^\circ} C $) and neutral gas far
from its critical point (for example argon). We consider the approximation
of a flow without viscosity. With an aim of simplifying the diagram of the
system of cooling of fast nuclear engines, we separate the coolant circuit
in two parts:\newline
- a first part of constant section $s $ and volume $v$, in which the flow is
carried along with cavitation. Cooling agent consists of sodium either in
liquid form or in form of vapour bubbles; argon is dissolved in the liquid
sodium or mixed with its vapour in the bubbles. The mixture contains $n $
bubbles each one of volume $v_1$. The bubbles are supposed in equilibrium in
a reference frame convected by the flow and the edge effects of the pipe are
not taken into account. The bubbles supposed without interactions are
identical and in spherical form.\newline
- a second part of constant section  $s_0 $ and volume $v_0 $ without
cavitation in which sodium is in liquid form and saturated by dissolved
argon.

\noindent The motion of the mixture in each compartment is supposed to be
uniform. One denotes respectively by $u $ and $u_0 $ the velocity values in
each compartment. The loop is closed and contains a total mass $M $ of fluid
(sodium). A pump discharges the fluid according to an imposed flow $d$. We
denote by $\rho_m $ the average density of the fluid in the cavitating part.
\newline
\noindent The theorem of Bernoulli is supposed to be applicable to each of
the two components of the fluid-gas mixture. This simplifying assumption
associated with a barotropic permanent motion makes it possible to connect
the velocity to the density of each component. The theorem of Bernoulli
extends to dynamics the equality of the chemical potentials between phases
and components [4]. These relations can be replaced by more advancing
equations of the motion representing, for a permanent flow, the connections
between velocities and densities.

\noindent The volumic free energy of the fluid-gas mixture is supposed to be
the sum of partial volumic free energies. The pressure of mixture is the sum
of the partial pressures. \bigskip The set of the assumptions leads to the
following relations:
\begin{equation*}
\rho_m v=\rho_l v+n v_1\left(\rho _v-\rho_l\right),
\end{equation*}
which gives the value of the fluid mass in the zone of cavitation; $\rho_l $
and $\rho _v $ stand for the densities of liquid and vapour. It follows:
\begin{equation*}
n={\frac{v}{v_1}} \big({\frac{\rho_l-\rho _m}{\rho_l-\rho _v}}\big). %
\eqno{(1)}
\end{equation*}
The total fluid mass in the coolant circuit is:
\begin{equation*}
M=\rho _mv+\rho _0v_0, \eqno{(2)}
\end{equation*}
where $\rho _0 $ denotes the density of the fluid in the part without
cavitation. \noindent A pump gives the fluid flow $d $ in the circuit,
\begin{equation*}
s\,\rho _mu=s_0\rho _0u_0=d . \eqno{(3)}
\end{equation*}
Let us denote by $F $ the volumic free energy of the fluid. The theorem of
Bernoulli applied to each phase of the fluid component gives:
\begin{equation*}
{\frac{1}{2}}u^2+F^{\prime }\left(\rho_l\right)={\frac{1}{2}}u^2_0+F^{\prime
}\left(\rho _0\right) \qquad \hbox{\rm (for the liquid phase),}\eqno{(4)}
\end{equation*}
\begin{equation*}
\ \ {\frac{1}{2}}u^2+F^{\prime }\left(\rho _v\right)={\frac{1}{2}}%
u^2_0+F^{\prime }\left(\rho _0\right)\qquad
\hbox{\rm (for the
vapour phase).}\eqno{(5)}
\end{equation*}
The approximations suppose a moderate cavitation. Although very synopsis,
this schematization allows to get a first estimate of the maximum number of
bubbles in the zone with cavitation. We take a state law of van der Waals
type for expressing the fluid partial pressure. The free volumic energy \ $%
F\ $ is given by a single expression in the liquid and the vapour phases
[5]. The relations (4) and (5) imply:
\begin{equation*}
F^{\prime }\left(\rho_l\right)=F^{\prime }\left(\rho _v\right) \eqno{(6)}
\end{equation*}
which is the equality of the chemical potentials of the two phases of the
fluid component. The free energy of the fluid being a non-convex function of
\ $\rho ,\ $ the values of \ $\rho_l\ $ and \ $\rho_v\ $ are different.
\newline
\noindent We denote by \ $G\ $ the volumic free energy of the gas. We obtain
$\ G^{\prime }\left(r_v\right)=G^{\prime }\left(r_l\right) $ where \ $r_v\ $
is the density of argon in the bubble and \ $r_l\ $ the density of saturated
gas dissolved in the liquid. Argon being a gas at a temperature far from its
critical value, the function \ $G\ $ is a convex function of the density \ $r
$. Consequently, $\ r_v=r_l $.

\noindent The partial pressures of the fluid and the gas are respectively:
\begin{equation*}
p_s\left(\rho \right)=\rho F^{\prime }\left(\rho \right)-F\left(\rho \right)%
\hbox{ \ \ et \ \ } p_a\left(r\right)=r\ G^{\prime }\left(r \right)-G\left(r
\right).
\end{equation*}
Within the framework of the Laplace model, the equilibrium of a bubble in
the zone of cavitation is:
\begin{equation*}
p_s(\rho _v)+p_a(r_v)-p_s(\rho _l)-p_a(r_l)={\frac{2\sigma }{R}}\,, %
\eqno{(7)}
\end{equation*}
where $\sigma $ is the constant surface tension at the temperature of the
flow and $R $ is the radius of the bubble [6]. Owing to the fact that $%
r_v=r_l $ and with $K$ denoting $\displaystyle
\left(36\pi\right)^{1/3}\sigma $, the relation (7) is written in the
equivalent form:
\begin{equation*}
F(\rho_l)-F(\rho _v)-F^{\prime }(\rho_l)(\rho_l-\rho _v)={\frac{2}{3}}%
Kv_1^{-1/3} .\eqno{(8)}
\end{equation*}
This relation remains unchanged if adding to $F $ an unspecified linear
function of $\rho $. Thus, $F^{\prime }(\rho ) $ is chosen null for the
densities of saturation $\rho _{ls} $ and $\rho _{vs} $ of the liquid and
the vapour. \newline
\noindent The volumic free energy is supposed to be a regular function near
the densities of saturation. The Taylor expansions to the second order in
the vicinity of \ $\rho _{ls}\ $ and \ $\rho _{vs} \ $ yield:
\begin{equation*}
F(\rho _v)={\frac{1}{2}}(\rho _v-\rho _{vs})^2F^{\prime \prime }(\rho_{vs} ),
\end{equation*}
\begin{equation*}
F(\rho_l)={\frac{1}{2}}(\rho_l-\rho _{ls})^2F^{\prime \prime }(\rho_{ls} ).
\end{equation*}

With taking into account the relation (6), we have
\begin{equation*}
F^{\prime \prime }(\rho _{vs})(\rho _v-\rho _{vs})=F^{\prime \prime }(\rho
_{ls})(\rho_l-\rho _{ls}),
\end{equation*}
and the relation (8) yields:
\begin{equation*}
F^{\prime \prime }(\rho _{ls})(\rho _{ls}-\rho_l)(\rho_l-\rho _v)\left\{ 1+{%
\frac{1}{2}}{\frac{\rho _{ls}-\rho_l}{\rho_l-\rho _v}} \left(1-{\frac{%
F^{\prime \prime }(\rho _{ls})}{F^{\prime \prime }(\rho _{vs})}}\right)
\right\}={\frac{2}{3}}Kv_1^{-1/3} . \eqno{(9)}
\end{equation*}
The value \ $C_l $ of the sound velocity in the liquid is
\begin{equation*}
C^2_l=\left({\frac{\partial p_{s}}{\partial \rho_l}}\right)\Bigg\vert%
_{\rho_{ls}} = \left(\rho F^{\prime }\left(\rho
\right)-F\left(\rho\right)\right)^{\prime }\Bigg\vert_{\rho _{ls}} =\rho
_{ls}F^{\prime \prime }\left(\rho_{ls}\right) .
\end{equation*}
The value \ $C_v $ of the sound velocity in the saturating vapour is
\begin{equation*}
C^2_v=\rho _{vs}F^{\prime \prime }\left(\rho _{vs}\right) .
\end{equation*}
So, we obtain
\begin{equation*}
\displaystyle {\frac{F^{\prime \prime }\left(\rho _{ls}\right) }{F^{\prime
\prime }(\rho _{vs})}}={\frac{C^2_l}{C^2_v}}{\frac{\rho _{vs}}{\rho _{ls}}} .
\end{equation*}
Far from the critical point of the fluid, we have
\begin{equation*}
{{\frac{F^{\prime \prime }(\rho _{ls} ) }{F^{\prime \prime }(\rho _{vs}) }}}
\ll 1 .
\end{equation*}
Such is the case of liquid sodium at the temperature of the nuclear engine
[7]. The range of values of $\rho _v $ and $\rho _l $ allows to write:
\begin{equation*}
{\frac{\rho _{ls}-\rho _l}{\rho _l-\rho _v}} \ll 1 .
\end{equation*}
So, from the respective values of the various terms of relation (9), one
deduces the approximated relation:
\begin{equation*}
v_1=\left( {\frac{2K}{3C^2_l}}\right)^3{\frac{1}{\left(\rho _{ls}-\rho _l
\right)^3}}\, . \eqno{(10)}
\end{equation*}

\section{An estimation of an upper bound value of the number of bubbles}

With the approximation (10), the relation (1) is written in the form:
\begin{equation*}
n={\frac{v}{v_{1}}}\,{\frac{(\rho _{l}-\rho _{m})}{\rho _{ls}}}\,.
\end{equation*}%
By taking into account relation (10), one obtains the number of bubbles in
the flow:
\begin{equation*}
n={\frac{3C_{l}^{6}}{32\pi \sigma ^{3}\rho _{ls}}}\ v\left( \rho _{ls}-\rho
_{l}\right) ^{3}\left( \rho _{l}-\rho _{m}\right) .\eqno{(11)}
\end{equation*}%
Let us notice that $\rho _{l}$ and $\rho _{m}$ are not independent but
bounded by the relations (2), (3) and (4) expressing $\rho _{m}$ as function
of $\rho _{l}$. Let us regard $n$ as a continuous quantity; from these
relations one deduces the differential system:
\begin{equation*}
\left\{
\begin{array}{l}
\quad {v\ d\rho _{m}+v_{0}\ d\rho _{0}=0}, \\
\quad {u\ d\rho _{m}+\rho _{m}\ du=0}, \\
\quad {u_{0}\ d\rho _{0}+\rho _{0}\ du_{0}=0}, \\
\quad {u\ du+F^{\prime \prime }(\rho _{l})d\rho _{l}-u_{0}\
du_{0}-F^{\prime
\prime }(\rho _{0})d\rho _{0}=0}.%
\end{array}%
\right.
\end{equation*}%
By eliminating \ $du,\ du_{0}\ $ et \ $d\rho _{0}\ $ between the four
relations one obtains:
\begin{equation*}
\left( {\frac{u^{2}}{\rho _{m}}}+{\frac{u_{0}^{2}v}{\rho _{0}v_{0}}}%
-F^{\prime \prime }(\rho _{0}){\frac{v}{v_{0}}}\right) d\rho _{m}=F^{\prime
\prime }(\rho _{l})d\rho _{l}\,.
\end{equation*}%
The respective orders of magnitude of the different terms of the preceding
relation and the fact that $\rho _{0}$ and $\rho _{l}$ are close to $\rho
_{ls}$ yield:
\begin{equation*}
-F^{\prime \prime }(\rho _{0}){\frac{v}{v_{0}}}\,d\rho _{m}=F^{\prime \prime
}(\rho _{l})d\rho _{l}\,.
\end{equation*}%
One obtains the derivative of $\rho _{m}$ as a function of $\rho _{l}$ in
the simplified form:
\begin{equation*}
{\frac{d\rho _{m}}{d\rho _{l}}}=-{\frac{v_{0}}{v}}\,.
\end{equation*}%
Relation (11) where $n$ is considered as a function only of $\rho _{m}$,
implies:
\begin{equation*}
n^{\prime }(\rho _{l})={\frac{3C_{l}^{6}}{32\pi \sigma ^{3}\rho _{ls}}}\
v\left\{ -3\left( \rho _{ls}-\rho _{l}\right) ^{2}\left( \rho _{l}-\rho
_{m}\right) +\left( \rho _{ls}-\rho _{l}\right) ^{3}\left( 1+{\frac{v_{0}}{v}%
}\right) \right\} .
\end{equation*}%
For $\rho _{l}$ different of $\rho _{ls},\,n^{\prime }(\rho _{l})$ is null
for:
\begin{equation*}
\rho _{l}-\rho _{m}={\frac{1}{3}}\left( \rho _{ls}-\rho _{l}\right) \left( 1+%
{\frac{v_{0}}{v}}\right) .\eqno{(12)}
\end{equation*}%
Let us notice that $n=0$ for $\rho _{l}=\rho _{ls}$ and $\rho _{l}=\rho _{m}$%
. The value of $\rho _{l}$ associated with relation (12) corresponds to a
maximum of bubbles in the flow. The value of the number of bubbles per unit
of volume is $\displaystyle N={\frac{n}{v}}\,$ such as:
\begin{equation*}
N={\frac{C_{l}^{6}}{32\pi \sigma ^{3}\rho _{ls}}}\left( \rho _{ls}-\rho
_{l}\right) ^{4}\left( 1+{\frac{v_{0}}{v}}\right) .\eqno{(13)}
\end{equation*}%
Instead of the density of liquid, it is more realistic to consider the flow
pressure in the part of the circuit subjected to cavitation. Indeed, this
physical quantity is more accessible to measurements. \newline
\noindent Let us denote by $\displaystyle\mathcal{P}=p_{s}(\rho
_{l})+p_{a}(r_{v})$ and $\displaystyle\mathcal{P}_{sat}=p_{s}(\rho
_{ls})+p_{a}(r_{v})$, where $p_{s}(\rho _{ls})$ is the saturated vapour
pressure and $p_{a}(r_{v})$ is the partial pressure of the gas. One obtains
\begin{equation*}
\mathcal{P}_{sat}-\mathcal{P}=C_{l}^{2}(\rho _{ls}-\rho _{l}).
\end{equation*}%
One deduces the value of the maximum number of bubbles per unit of volume in
the form:
\begin{equation*}
N={\frac{\left( \mathcal{P}_{sat}-\mathcal{P}\right) ^{4}}{32\pi \sigma
^{3}C_{l}^{2}\rho _{ls}}}\ \left( 1+{\frac{v_{0}}{v}}\right) .\eqno{(14)}
\end{equation*}

For example, we consider the physical values of liquid sodium at the
temperature of $400 {\ ^\circ} C$: surface tension $\sigma =0,166\ N/m\ $;
sound velocity liquid sodium $\displaystyle C_l=2350\ m/s\ $; density of
saturating sodium $\displaystyle \rho _{ls}=856\ Kg/m^3\ $ [7].

With choosing for the coefficient $\displaystyle
1+{\frac{v_0}{v}}\, $ the value 10 and for difference of pressures $\mathcal{%
P}_{sat} - \mathcal{P} = 4000 \ Pa $, one obtains in the part prone to
cavitation a maximum of bubbles $N = 1200 $ per $liter$ .

\section{Conclusion}

\quad We presented an approach allowing to consider a higher limit
number of the density of bubbles present in a closed loop of
cavitating flow. This number is given by the relation (14). The
effects of uncondensable gases (argon in our case) do not appear in
the relation owing to the fact that the fluid (sodium) is saturated
with gas. The bubbles are supposed in local equilibrium, which
requires that their times of creation and collapse are weak with
respect to the running time in the cavitating part of the circuit (a
fine study could be carried out starting from the equation of
Rayleigh-Plesset [8]).\newline \quad The relation binding the
average densities of the fluid $\rho_m $ and of the liquid $\rho_l $
partially results from the relation (4). This relation corresponds
to a permanent flow. It may be replaced by all other relations
binding $u, u_0, \rho_l$ et $\rho_0$. \newline \quad The method
corresponds to a very simplified model of a flow of complex nature.
Nevertheless, this model offers the advantage of showing that in a
closed pipe, cavitation is a limited phenomenon. It should be noted
that relation (14) takes into account the ratio of dimensions
between the zones of cavitation and without cavitation. To decrease
cavitation, one may reduce this ratio by creating additional zones
where cavitation is of no importance on the maintenance of the
coolant circuit.

\vfill\eject

\centerline {\bf Abridged French version} \vskip 0.1cm {\small
\centerline {\bf R\'esum\'e} }

{\small La cavitation est un ph\'enom\`ene g\'en\'eral des
\'ecoulements fluides avec obstacles. Elle apparait dans les
conduites de refroidissement des r\'eacteurs nucl\'eaires rapides.
Un mod\`ele de ce ph\'enom\`ene utilisant la th\'eorie de Laplace et
une \'energie non convexe commune aux phases liquide et vapeur est
propos\'e. Il permet de d\'eterminer une borne sup\'erieure de la
densit\'e de bulles (nombre de bulles par unit\'e de volume dans
l'\'ecoulement). L'intensit\'e maximum de la cavitation est
associ\'ee aux caract\'eristiques m\'ecaniques et thermiques de
l'\'ecoulement fluide.}\newline

{\small \centerline{\bf Version fran\c caise  abr\'eg\'ee} }

{\small Il a \'et\'e constat\'e exp\'erimentalement que le nombre de
bulles dans un \'ecoulement cavitant d\'epend de la qualit\'e du
fluide. En particulier, le nombre de particules et de bulles
microscopiques de gaz est un param\`etre qu'il est important de
conna\^itre [1-2]. Malheureusement, ce nombre est difficilement
mesurable dans la plupart des \'ecoulements. C'est notamment le cas
pour l'\'ecoulement de sodium liquide en r\'eacteur nucl\'eaire de
type RNR. D'autres exp\'eriences ont r\'ev\'el\'e l'existence, pour
un d\'ebit fix\'e, d'une valeur maximale pour le nombre de bulles;
au del\`a de ce seuil, l'injection de microbulles de gaz ne semble
plus modifier l'intensit\'e de la cavitation [3]. \newline
L'objectif de ce travail est de quantifier l'intensit\'e de la
cavitation \`a partir de l'estimation analytique du nombre de bulles
dans un \'ecoulement. Nous consid\'erons un \'ecoulement diphasique
permanent isotherme constitu\'e d'un "fluide" (par exemple du sodium
\`a la temp\'erature de $400 {\ ^\circ} C$) et d'un "gaz"  neutre
tr\`es au-dessus de son point critique (par exemple de l'argon).
Nous envisageons l'approximation d'un \'ecoulement sans viscosit\'e.
Dans le but de simplifier le sch\'ema du syst\`eme de
refroidissement des r\'eacteurs nucl\'eaires rapides, nous
s\'eparons le circuit de refroidissement en deux parties: \newline
\quad - Une premi\`ere partie de section constante $s $ et de volume
$v$, dans laquelle l'\'ecoulement s'effectue avec cavitation. Le
r\'efrig\'erant est constitu\'e de sodium soit sous forme liquide
soit sous forme de bulles de vapeur; l'argon est dissous dans le
sodium liquide ou m\'elang\'e \`a sa vapeur dans les bulles. Le
m\'elange contient $n $ bulles, chacune de volume $v_1. $ Les bulles
sont suppos\'ees en \'equilibre dans un rep\`ere entra\^in\'e par
l'\'ecoulement et les effets de bord de la conduite ne sont pas pris
en compte. Les bulles suppos\'ees sans interactions entre elles sont
alors identiques et d'une forme sph\'erique. La zone de cavitation
correspond \`a la r\'egion d'\'echange de chaleur entre les barres
radioactives et le fluide r\'efrig\'erant. \newline \quad - Une
deuxi\`eme partie suppos\'ee sans cavitation, de section constante
$s_0 $ et de volume   $v_0 $ dans laquelle le sodium est sous la
forme liquide et satur\'e par de l'argon dissous.
\newline \noindent Le mouvement du m\'elange dans chaque
compartiment est pris uniforme. On note respectivement par   $u  $
et   $u_0  $ les valeurs de la vitesse dans chaque compartiment. Le
circuit est ferm\'e et contient une masse totale donn\'ee   $M  $ de
fluide. Une pompe entra\^ine le fluide suivant un d\'ebit impos\'e
$d$. Nous noterons   $\rho_m  $ la masse volumique moyenne du fluide
dans la partie cavitante. \newline Le th\'eor\`eme de Bernoulli est
suppos\'e applicable \`a chacun des deux constituants du m\'elange
fluide-gaz. Cette hypoth\`ese simplificatrice associ\'ee \`a un
mouvement permanent et barotrope permet de relier la vitesse \`a la
masse volumique de chaque constituant. Ces relations peuvent \^etre
remplac\'ees par des \'equations du mouvement plus \'evolu\'ees
repr\'esentant, lorsque l'\'ecoulement est permanent,  les liaisons
entre la vitesse et la masse volumique de chaque constituant.
\newline Consid\'erons pour simplifier des bulles en \'equilibre
dans un rep\`ere li\'e \`a cet \'ecoulement. En utilisant la
th\'eorie de Laplace, l'\'energie libre totale du milieu s'\'ecrit
comme la somme des \'energies de chaque phase et de l'\'energie
superficielle des bulles. On obtient alors l'\'egalit\'e des
potentiels chimiques pour le constituant fluide et pour le gaz ainsi
que l'expression du saut de pression \`a l'interface. Ces relations
d\'eterminent les caract\'eristiques g\'eom\'etriques des bulles de
vapeur. On repr\'esente l'\'energie libre du fluide suppos\'ee non
convexe par deux paraboles dont les rayons de courbure aux sommets
sont reli\'es \`a la vitesse du son dans chaque phase. \newline
\noindent L'ensemble des hypoth\`eses effectu\'ees conduit \`a
l'expression (1) du nombre de bulles dans la zone de cavitation en
fonction des densit\'es du liquide et vapeur et de leur dimension,
\`a la conservation de la masse totale de fluide (relation (2)), \`a
l'expression du d\'ebit impos\'e (relation (3)). Le th\'eor\`eme de
Bernoulli appliqu\'e \`a chacune des phases du fluide conduit aux
relations (4) et (5) d'o\`u on d\'eduit l'\'egalit\'e (6) des
potentiels chimiques [4]. De la d\'efinition de la pression \`a
partir de l'\'energie libre et de la loi de Laplace on obtient
l'expression (8) reliant le rayon des bulles aux densit\'es du
liquide et de sa vapeur. En se limitant aux termes principaux, on
montre que les deux relations (6) et (8) se r\'eduisent alors \`a
une seule expression: l'\'equation (10) reliant le volume des bulles
\`a la densit\'e du liquide dans la zone cavitante $\rho_l$. La
pr\'esence de gaz qui sature le fluide ne modifie pas la forme de
l'\'equation de Laplace. \newline Du fait des relations (2), (3) et
(4), le nombre de bulles $n$ peut \^etre consid\'er\'e comme un
fonction uniquement de $\rho_l$. On constate alors que le nombre $n$
de bulles admet une valeur maximale. Cette valeur prend une forme
explicitable en fonction de l'\'ecart entre la densit\'e du liquide
et sa densit\'e \`a la pression de saturation. C'est l'expression
(13) donnant le nombre maximum de bulles par unit\'e de volume. Elle
constitue donc une premi\`ere majoration analytique du nombre de
bulles dans un \'ecoulement cavitant. \newline La densit\'e du
sodium liquide n'\'etant pas une grandeur physiquement mesurable, il
convient de transformer la relation (13). En utilisant une loi
d'\'etat lin\'eaire, on exprime le nombre maximum de bulles non plus
en fonction de la densit\'e du sodium liquide dans la zone de
cavitation, mais en fonction de la pression totale. La relation
obtenue (14) montre que le nombre maximum de bulles par unit\'e de
volume est fonction de l'\'ecart entre la pression totale et la
pression totale \`a la saturation. Un calcul num\'erique est
effectu\'e pour du sodium \`a la temp\'erature des circuits de
refroidissement des r\'eacteurs nucl\'eaires. Il constitue un ordre
de grandeur r\'ealiste du nombre de bulles d\`es lors que l'on est
capable de mesurer la pression dans la zone cavitante. }

\end{document}